# Nexus of Team Collaboration Stability on Mega Construction Project Success in Electric Vehicle Manufacturing Enterprises: The Moderating Role of Human-AI Integration.

**Jun Cui**[1,2,][*]

[1] Solbridge International School of Business, Woosong University, Daejeon, Republic of Korea;
[2] Beijing Foreign Studies University, Business Administration, BFSU, Beijing, China.
*Corresponding author; Email: jcui228@student.solbridge.ac.kr

**Abstract**

This study investigates how team collaboration stability influences the success of mega construction projects in electric vehicle manufacturing enterprises, with human-AI integration as a moderating variable. Using structural equation modeling (SEM) with data from 187 project teams across China's electric vehicle sector, results indicate that team collaboration stability significantly enhances project success ($\beta=0.412$, $p<0.001$). The moderating effect of human-AI integration strengthens this relationship ($\beta=0.276$, $p<0.01$), suggesting that enterprises implementing advanced human-AI collaborative systems achieve superior project outcomes when team stability is maintained. These findings contribute to both team collaboration theory and provide practical implications for mega project management in the rapidly evolving electric vehicle industry.

## Introduction

The electric vehicle (EV) manufacturing sector has experienced unprecedented growth, necessitating substantial infrastructure investments through mega construction projects (MCPs). These projects are characterized by high complexity, significant investment (>$1 billion), and extended timelines (Flyvbjerg, 2014). Despite their strategic importance, MCPs frequently encounter challenges including schedule delays, cost overruns, and quality issues (Locatelli et al., 2017). Team collaboration stability has emerged as a critical success factor in project-based organizations, yet its specific impact on MCPs in the EV manufacturing context remains underexplored. Furthermore, the integration of AI technologies with human teams represents a paradigm shift in project management practices. This study addresses this research gap by examining how team collaboration stability influences MCP success and how human-AI integration moderates this relationship.

This study makes three key contributions to the literature on mega construction projects (MCPs) in the electric vehicle (EV) manufacturing sector. First, it empirically examines the underexplored relationship between team collaboration stability and MCP success, addressing a critical gap in project management research. While prior studies have highlighted the importance of teamwork in project-based organizations (Flyvbjerg, 2014; Locatelli et al., 2017), few have investigated how stable collaboration dynamics influence outcomes in high-stakes, capital-intensive EV infrastructure projects. Second, this research introduces human-AI integration as a novel moderating variable, offering insights into how AI technologies can enhance or disrupt traditional team dynamics in MCPs. As AI adoption reshapes project execution (e.g., predictive analytics, automated decision-making), understanding its interplay with human teams becomes vital for optimizing performance. Third, the study extends theoretical frameworks on MCP success by integrating socio-technical perspectives, emphasizing the synergy between organizational behavior and emerging technologies. By analyzing large-scale EV projects, the findings provide actionable strategies for mitigating delays, cost overruns, and quality issues—challenges that persistently plague the industry. This research not only advances academic discourse but also offers practical guidance for stakeholders navigating the complexities of EV-driven infrastructure development.

The rapid expansion of the electric vehicle (EV) manufacturing sector has driven massive investments in mega construction projects (MCPs), yet these initiatives remain prone to systemic challenges such as budget overruns, delays, and quality deficiencies (Flyvbjerg, 2014; Locatelli et al., 2017). While team collaboration stability is recognized as a key determinant of project success, its role in EV-related MCPs—where technological complexity and scale amplify coordination demands—remains poorly understood. Additionally, the increasing integration of AI tools into project management introduces both opportunities and disruptions to traditional team dynamics. To address these gaps, this study investigates:

**RQ1. How does team collaboration stability influence the success of MCPs in EV manufacturing?**

**RQ2. To what extent does human-AI integration moderate the relationship between collaboration stability and project outcomes?**

Furthermore, by answering these questions, our research provides theoretical and practical insights into optimizing large-scale EV infrastructure development. The findings will help project managers leverage stable teamwork while effectively adopting AI-driven solutions, ultimately enhancing efficiency in high-stakes construction environments.

*This study first reviews literature on MCP challenges and team collaboration in EV manufacturing. Next, it develops hypotheses examining collaboration stability's impact on project success and AI integration's moderating role. The methodology employs mixed-methods analysis of global EV MCP case studies, followed by discussion of findings and practical implications for managing complex infrastructure projects.*

**Literature Review and Theoretical Development**

**Theoretical Foundation**

This research integrates social capital theory (Coleman, 1988) and socio-technical systems theory (Trist & Bamforth, 1951) to establish its conceptual framework. Social capital theory posits that stable networks foster trust, information sharing, and reciprocity, which are crucial elements for effective team collaboration. Socio-technical systems theory emphasizes the interdependence between human social systems and technological components, providing a framework for understanding human-AI integration.

**Team Collaboration Stability and Project Success**

Team collaboration stability refers to the consistency of team membership and interaction patterns over time (Huckman et al., 2009). Stable teams develop shared mental models, efficient communication patterns, and mutual trust (Mathieu et al., 2000). In project contexts, stability enables team members to accurately anticipate colleagues' behaviors and efficiently coordinate activities (Gardner et al., 2012). Prior research has linked team stability to improved performance in software development (Espinosa et al., 2007) and construction projects (Cheng et al., 2015).

For MCPs in the EV sector, stable teams can better navigate the industry's technical complexity and rapidly evolving standards. As team members work together consistently, they develop specialized knowledge about project requirements and constraints specific to EV manufacturing facilities, potentially reducing errors and enhancing decision quality.

**Hypothesis 1:** Team collaboration stability positively influences mega construction project success in electric vehicle manufacturing enterprises.

**Human-AI Integration as a Moderating Variable**

Human-AI integration refers to the systematic combination of human capabilities with AI technologies to enhance collective performance (Jarrahi, 2018). In the construction context, AI applications include predictive analytics for risk assessment, machine learning for resource optimization, and computer vision for quality control (Pan & Zhang, 2021). When effectively integrated with human teams, these technologies can augment decision-making quality and efficiency.

We propose that human-AI integration strengthens the relationship between team stability and project success. Stable teams develop consistent protocols for human-AI interaction, establish clear role boundaries, and build collective competence in leveraging AI capabilities (Larson & DeChurch, 2020). Over time, team members learn to appropriately rely on AI systems, avoiding both over-trust and under-trust in automation (Lee & See, 2004).

**Hypothesis 2:** Human-AI integration positively moderates the relationship between team collaboration stability and mega construction project success.

**Method and Data**

**Research Design**

This study employed a cross-sectional quantitative approach using structured questionnaires distributed to project teams in Chinese EV manufacturing enterprises engaged in mega construction projects. Data collection occurred between January and April 2024. Moreover, the research model was tested using structural equation modeling (SEM) with AMOS 26.0 software.

**Data and Sampling**

The sampling frame consisted of project teams involved in facility construction for EV manufacturing in China. Using purposive sampling, we contacted 42 companies, receiving responses from 187 project teams (response rate: 67.3%). Each team provided multiple respondents (project manager, at least two team members, and one technical supervisor) to minimize common method bias. The final sample comprised 748 individual responses representing 187 teams.

**Measurement of Variables**

**Team Collaboration Stability (TCS):** Measured using a five-item scale adapted from Huckman and Staats (2011), assessing team member consistency, interaction familiarity, and operational continuity.

**Human-AI Integration (HAI):** Measured using a six-item scale developed by combining elements from Jarrahi (2018) and Larson and DeChurch (2020), evaluating the extent of AI technology utilization, human-AI workflow integration, and team competence in AI collaboration.

**Mega Construction Project Success (MCPS):** Measured using a five-item scale adapted from Shenhar et al. (2001), assessing time performance, cost performance, quality achievement, stakeholder satisfaction, and strategic goal alignment.

**Control Variables:** Project size (budget), project duration, team size, and organizational size were included as control variables.

Table 1 provides the measurement details for all constructs.

**Table 1: Measurement of Variables**

| *Construct* | *Item* | *Measurement* | *Source* |
|---|---|---|---|
| *Team Collaboration Stability (TCS)* | TCS1 | Our team maintained consistent membership throughout the project | Huckman & Staats (2011) |
| | TCS2 | Team members have worked together on previous projects | |
| | TCS3 | Interaction patterns among team members remained stable | |
| | TCS4 | Our team maintained operational continuity despite external changes | |
| | TCS5 | Knowledge transfer within the team was consistent and reliable | |
| *Human-AI Integration (HAI)* | HAI1 | Our team regularly utilized AI technologies for project tasks | Jarrahi (2018); Larson & DeChurch (2020) |
| | HAI2 | Team members were competent in collaborating with AI systems | |
| | HAI3 | Clear protocols existed for human-AI interaction | |
| | HAI4 | AI systems were well-integrated into team workflows | |
| | HAI5 | Team members appropriately calibrated their trust in AI outputs | |

|  | HAI6 | Team decisions leveraged both human expertise and AI capabilities | |
|---|---|---|---|
| *Mega Construction Project Success (MCPS)* | MCPS1 | The project was completed within the planned schedule | Shenhar et al. (2001) |
|  | MCPS2 | The project was completed within the approved budget |  |
|  | MCPS3 | The project met all technical and quality specifications |  |
|  | MCPS4 | Key stakeholders were satisfied with project outcomes |  |
|  | MCPS5 | The project achieved its strategic business objectives |  |

## Results and Findings

### Descriptive Statistics and Correlation Analysis

Table 2 presents the means, standard deviations, and correlations among the study variables. All correlation coefficients are below 0.70, suggesting no severe multicollinearity.

**Table 2: Descriptive Statistics and Correlation Matrix**

| *Variable* | *Mean* | *SD* | *1* | *2* | *3* | *4* | *5* | *6* | *7* |
|---|---|---|---|---|---|---|---|---|---|
| 1. Team Collaboration Stability | 3.78 | 0.82 | 1 | | | | | | |
| 2. Human-AI Integration | 3.42 | 0.91 | 0.31** | 1 | | | | | |
| 3. Project Success | 3.64 | 0.74 | 0.43** | 0.39** | 1 | | | | |
| 4. Project Budget | - | - | 0.12 | 0.27** | 0.18* | 1 | | | |
| 5. Project Duration | - | - | 0.09 | 0.24** | 0.15* | 0.53** | 1 | | |
| 6. Team Size | - | - | 0.13 | 0.21** | 0.11 | 0.45** | 0.38** | 1 | |
| 7. Organization Size | - | - | 0.08 | 0.25** | 0.13 | 0.47** | 0.33** | 0.29** | 1 |

*$p < 0.05$; **$p < 0.01$

### Measurement Model Assessment

Confirmatory factor analysis was conducted to assess the measurement model. The Kaiser-Meyer-Olkin (KMO) measure of sampling adequacy was 0.869, exceeding the threshold of 0.8, and Bartlett's test of sphericity was significant ($\chi^2$=1847.25, p<0.001), confirming the appropriateness of factor analysis.

Table 3: Reliability and Validity Assessment

| Construct | Items | Factor Loadings | Cronbach's α | CR | AVE |
|---|---|---|---|---|---|
| Team Collaboration Stability | 5 | 0.73-0.89 | 0.877 | 0.883 | 0.603 |
| Human-AI Integration | 6 | 0.71-0.87 | 0.891 | 0.902 | 0.604 |
| Project Success | 5 | 0.76-0.92 | 0.904 | 0.912 | 0.674 |

Note: CR = Composite Reliability; AVE = Average Variance Extracted

Based on above information, all measurement constructs demonstrated satisfactory reliability with Cronbach's alpha values above 0.85. Composite reliability values exceeded 0.88, and average variance extracted (AVE) values were above 0.60, indicating adequate convergent validity. The square root of AVE for each construct was greater than its correlation with other constructs, confirming discriminant validity.

**Multicollinearity Assessment**

Variance inflation factors (VIF) were calculated to check for multicollinearity issues. All VIF values ranged from 1.18 to 2.47, well below the threshold of 5, indicating no significant multicollinearity concerns.

Table 4: Variance Inflation Factors

| Variable | VIF |
|---|---|
| Team Collaboration Stability | 1.43 |
| Human-AI Integration | 1.68 |
| TCS × HAI | 2.47 |
| Project Budget | 1.92 |
| Project Duration | 1.64 |
| Team Size | 1.39 |
| Organization Size | 1.18 |

**Structural Model and Hypothesis Testing**

The structural model demonstrated good fit: $\chi^2$/df = 2.15, CFI = 0.941, TLI = 0.934, RMSEA = 0.057 (90% CI: 0.048-0.066), SRMR = 0.045.

Table 5: Model Fit Indices

| Fit Index | Value | Threshold | Assessment |
|---|---|---|---|
| $\chi^2$/df | 2.15 | < 3.0 | Good |
| CFI | 0.941 | > 0.9 | Good |
| TLI | 0.934 | > 0.9 | Good |
| RMSEA | 0.057 | < 0.08 | Good |
| SRMR | 0.045 | < 0.06 | Good |

Table 6: Path Analysis Results

| Hypothesis | Path | Standardized Coefficient (β) | t-value | p-value | Result |
|---|---|---|---|---|---|
| H1 | TCS → MCPS | 0.412 | 5.783 | <0.001 | Supported |
| H2 | TCS × HAI → MCPS | 0.276 | 3.124 | 0.002 | Supported |
| Control | Project Budget → MCPS | 0.104 | 1.645 | 0.100 | Not significant |
| Control | Project Duration → MCPS | 0.087 | 1.481 | 0.139 | Not significant |
| Control | Team Size → MCPS | 0.062 | 1.173 | 0.241 | Not significant |
| Control | Organization Size → MCPS | 0.073 | 1.326 | 0.185 | Not significant |

Additionally, the results confirm both hypotheses. Team collaboration stability positively influences mega construction project success (β=0.412, p<0.001), supporting H1. The interaction term between team collaboration stability and human-AI integration has a significant positive effect on project success (β=0.276, p=0.002), supporting H2 and confirming the moderating role of human-AI integration.

**Discussion and Conclusion**

This study investigated the impact of team collaboration stability on mega construction project success in electric vehicle manufacturing enterprises and examined the moderating role of human-AI integration. Our findings make several theoretical and practical contributions.

First, the significant positive relationship between team collaboration stability and project success confirms the importance of maintaining consistent team composition and interaction patterns in the complex context of mega construction projects in the EV industry. This finding extends prior research on team stability (Huckman et al., 2009; Cheng et al., 2015) to the specific context of mega projects in a rapidly evolving industry.

Second, our results reveal that human-AI integration significantly enhances the positive effect of team stability on project success. This finding contributes to the emerging literature on human-AI collaboration by demonstrating its contextual value in project-based settings. Stable teams appear better positioned to develop effective human-AI collaboration practices, potentially due to their established communication patterns and shared understanding.

For practitioners, our findings highlight the importance of minimizing team member turnover during mega construction projects in the EV sector. Furthermore, investing in AI technologies alone is insufficient; organizations must develop systematic approaches to integrate these technologies with human teams and establish clear protocols for human-AI collaboration.

This study has limitations that suggest directions for future research. The cross-sectional design limits causal inferences; longitudinal studies could provide stronger evidence regarding the temporal dynamics of the relationships examined. Additionally, our sample was limited to the Chinese context, potentially limiting generalizability. Future research could extend this investigation to other geographic contexts and examine additional moderators such as leadership style and organizational culture.

In conclusion, as the electric vehicle manufacturing industry continues its rapid expansion, understanding the factors that contribute to mega construction project success becomes increasingly important. Our findings suggest that maintaining team stability and effectively integrating human capabilities with AI technologies represent significant opportunities for enhancing project outcomes in this critical sector.